\documentstyle[pra,aps,epsfig,amssymb]{revtex}

\begin{document}
%\psdraft
%%%%%%%%%%%%%%%%
% sp definitions:
\newcommand{\beq}{\begin{equation}}\newcommand{\eeq}{\end{equation}}
\newcommand{\barr}{\begin{eqnarray}}\newcommand{\earr}{\end{eqnarray}}

\newcommand{\andy}[1]{ }

\def\txt{\textstyle}
%%%%%%%%%%%%%%%%%%
% lss definitions:
\def\NI{\noindent} \def\lslittlenote#1{\lsnote{{\sl #1}}}
 \def\lsnote#1{\def\dash{\hbox{\rm---}}{\bf~[[}~{\tt #1}~{\bf]]~}}
\def\eqn#1{Eq.\ (\ref{eq:#1})} \def\pp{{\vec p}}
%\font\romsix=cmr6 scaled\magstep0
%%%%%%%%%%%%%%%%%
\def\ask{\marginpar{?? ask: \hfill}}
\def\fin{\marginpar{fill in ... \hfill}}
\def\note{\marginpar{note \hfill}}
\def\check{\marginpar{check \hfill}}
\def\discuss{\marginpar{discuss \hfill}}
%%%%%%%%%%%%%%%%%%
%%%%%%%%%%%%%%%%%%%%%%%%Nuove def%%%%%%%%%%%%%%%%%%%%%%%%%%%
\def\hh{\widehat}
\def\wtilde{\widetilde}
\newcommand{\bm}[1]{\mbox{\boldmath $#1$}}
\newcommand{\bmsub}[1]{\mbox{\boldmath\scriptsize $#1$}}
\newcommand{\bmh}[1]{\mbox{\boldmath $\hat{#1}$}}
\newcommand{\ket}[1]{| #1 \rangle}
\newcommand{\bra}[1]{\langle #1 |}
%%%%%%%%%%%%%%%%%%%%%%%%%%%%%%%%%%%%%%%%%%%%%%%%%%%%%%%%%%%

\begin{titlepage}
\begin{flushright}
\today \\
BA-TH/00-390\\
\end{flushright}
\vspace{.5cm}
\begin{center}
{\LARGE Decoherence vs entropy in neutron interferometry

}

\quad

{\large P. Facchi,\(^{1}\) A. Mariano\(^{2}\)  and S.
Pascazio\(^{2}\)
\\
           \quad \\
$^{1}$Atominstitut der \"Osterreichischen Universit\"aten,
Stadionallee
2, A-1020, Wien, Austria \\
$^2$Dipartimento di Fisica, Universit\`a di Bari \\
     and Istituto Nazionale di Fisica Nucleare, Sezione di Bari \\
 I-70126 Bari, Italy \\

}

\vspace*{.5cm} PACS: 03.65.Bz, 03.75.Be, 03.75.Dg
\vspace*{.5cm}

{\small\bf Abstract}\\ \end{center}

{\small We analyze the coherence properties of polarized neutrons,
after they have interacted with a magnetic field or a phase
shifter undergoing different kinds of statistical fluctuations.
We endeavor to probe the degree of disorder of the distribution
of the phase shifts by means of the loss of quantum mechanical
coherence of the neutron. We find that the notion of entropy of
the shifts and that of decoherence of the neutron do not
necessarily agree. In some cases the neutron wave function is
more coherent, even though it has interacted with a more
disordered medium. }

\end{titlepage}

\newpage

\section{Introduction}

The notion of decoherence has attracted increasing attention in
the literature of the last few years \cite{Dec,NPN}. The loss of
quantum mechanical coherence undergone by a quantum system, as a
consequence of its interaction with a given environment, can be
discussed in relation to many different physical phenomena and
has deepened our comprehension of fundamental issues, disclosing
unexpected applications as well as innovative technology.

Neutron physics (neutron optics in particular) has played an
important role in this context, both on theoretical and
experimental grounds. Non-classical states are readily obtained,
for instance by splitting and then superposing wavepackets in an
interferometer \cite{neutron} or different spin states in a
magnetic field \cite{longSG,echo}, and are of great significance
in the investigation of fundamental quantum mechanical
properties. The aim of this paper is to investigate the coherence
features of neutron wave packets, by making use of the Wigner
function \cite{Wigner}, in analogy with concepts and techniques
that are routinely used in quantum optics \cite{QOpt}. The studies
of the last few years have shown that non-classical states are
fragile against statistical fluctuations
\cite{NP1,RS}: the analysis of situations where these states
display robustness during the interaction with noisy environments
is therefore of great practical interest.

The main motivation
of this work is to use the coherence properties of the wave
function as a ``probe" to check the degree of disorder of an
environment. A similar idea was first proposed, as far as we know,
in the context of quantum chaos and Feynman integrals
\cite{Saito}. One might naively expect that a neutron ensemble
suffers a greater loss of quantum coherence by interacting with
an increasingly disordered environment: intuitively, a more
disordered environment should provoke more randomization of the
phase of the wave function, which in turn implies more quantum
decoherence. As we shall see, this is not always true: some of
the results to be discussed below are rather counterintuitive and
at variance with naive expectation. In some cases the neutron
wave function is {\em more coherent}, even though it has
interacted with a {\em more disordered} medium. This statement
can be given a precise quantitative meaning in terms of the
entropy of the medium and of a ``decoherence parameter" that will
be defined for the neutron density matrix.

\section{Preliminaries }
\label{sec-prel}
\andy{prel}

The Wigner quasidistribution function \cite{Wigner} can be defined in
terms of the density matrix $\rho$ as
\andy{wigrho}
\beq\label{eq:wigrho}
W(x,k) = \frac{1}{2\pi}\int d\xi\; e^{-ik\xi}
\langle x+\xi/2|\rho|x-\xi/2\rangle ,
\eeq
where $x$ and $p=\hbar k$ are the position and momentum of the
particle. One easily checks that the Wigner function is
normalized to unity and its marginals represent the position and
momentum distributions
\andy{normW,margx,p}
\barr
& & {\rm Tr}\rho= \int dx\; dk\; W(x,k) =1,
\label{eq:normW}\\
& & P(x) = \bra{x}\rho\ket{x}= \int dk\; W(x,k), \label{eq:margx} \\
& & P(k) = \bra{k}\rho\ket{k}
=\int dx\; W(x,k).
\label{eq:margp}
\earr
 Notice that
\andy{tracesq}
\beq\label{eq:tracesq}
\int dx\;dk\;W(x,k)^2=\frac{{\rm Tr}\rho^2}{2\pi}.
\eeq
In this paper we shall consider a one-dimensional system (the
extension to 3 dimensions is straightforward) and assume that the
wave function is well approximated by a Gaussian
\andy{gauss,gaussinv}
\barr
\psi(x) &=& \bra{x}\psi\rangle= \frac{1}{(2\pi\delta^2)^{1/4}} \exp
\left[-\frac{(x-x_0)^2}{4\delta^2} + i k_0 x\right] ,
\label{eq:gauss} \\
\phi(k) &=& \bra{k}\psi\rangle=\frac{1}{(2\pi\delta_k^2)^{1/4}} \exp
\left[-\frac{(k-k_0)^2}{4\delta_k^2} - i(k-k_0)x_0\right]
\nonumber \\
 &=&
\left(\frac{2\delta^2}{\pi} \right)^{1/4}
\exp \left[-\delta^2(k-k_0)^2 - i(k-k_0)x_0\right],
\label{eq:gaussinv}
\earr
where $\psi(x)$ and $\phi(k)$ are the wave functions in the
position and momentum representation, respectively, $\delta$ is
the spatial spread of the wave packet, $\delta_k
\delta= 1/2$, $x_0$ is the initial average position of the particle
and $p_0=\hbar k_0$ its average momentum. The two functions above
are both normalized to one: normalization will play an important
role in our analysis and will never be neglected.
The Wigner function for the state
(\ref{eq:gauss})-(\ref{eq:gaussinv}) is readily calculated
\andy{wiggauss}
\beq\label{eq:wiggauss}
W(x,k) = \frac{1}{\pi}
\exp \left[-\frac{(x-x_0)^2}{2\delta^2}\right]
\exp \left[-2\delta^2(k-k_0)^2\right].
\eeq

In this paper we will focus on two physical situations. In the first
one, a polarized neutron acquires a phase shift $\Delta$, either by
going through a phase shifter or by crossing a magnetic field
parallel to its spin. In the second one, a polarized neutron is
divided in two states, either in an interferometer or by crossing a
magnetic field perpendicular to its spin. The latter situation is
physically most interesting, for it yields non-classical states,
whose coherence properties are of great interest.

\subsection{Single Gaussian}
\label{sec-sGauss}
\andy{sGauss}

If a Gaussian wave packet undergoes a phase shift $\Delta$, the
resulting Wigner function reads
\andy{wwigner}
\beq\label{eq:wwigner}
W(x,k,\Delta) = \frac{1}{\pi} \exp
\left[-\frac{(x-x_0+\Delta)^2}{2\delta^2}\right] \exp
\left[-2\delta^2(k-k_0)^2\right].
\eeq
Physically, this is achieved either by placing a phase shifter in
the neutron path, or by injecting a polarized neutron in a
constant magnetic field parallel to its spin. In both cases, the
total energy of the neutron is conserved. In the latter case, if
the field has intensity $B$ and is contained in a region of
length $L$, the neutron kinetic energy in the field changes by
$\Delta E = -|\mu| B$, where $|\mu|$ is the neutron magnetic
moment. This entails a change in average momentum $\Delta k= m\mu
B /\hbar^2 k_0$ and a phase shift proportional to $\Delta \equiv
L\Delta k/k_0$. When it leaves the field, the neutron acquires
again the initial kinetic energy.

\subsection{Double Gaussian}
\label{sec-dGauss}
\andy{dGauss}

Consider now a neutron wave packet that is split and then
recombined in an interferometer, with a phase shifter placed in
one of the two routes. The Wigner function in the ordinary
channel (transmitted component) is readily computed:
\andy{intcats}
\barr
W^{\rm int}(x,k,\Delta)&=&\frac{1}{4\pi}\exp[-2\delta^2(k-k_0)^2]
\left[\exp\left(-\frac{\left(x-x_0+\Delta\right)^2}{2\delta^2}\right)
\right. \nonumber\\ & & \left.
+\exp\left(-\frac{\left(x-x_0\right)^2}{2\delta^2}\right) +2
\exp\left(-\frac{\left(x-x_0+\frac{\Delta}{2}\right)^2}
{2\delta^2}\right)\cos(k\Delta)\right].
\label{eq:intcats}
\earr
Notice that, for $\Delta\neq0$, it is not normalized to unity
(some neutrons end up in the extraordinary channel---reflected
component) and that for $\Delta=0$ (no phase shifter) one
recovers (\ref{eq:wiggauss}).

A similar result is obtained when a polarized (say, $+y$) neutron
crosses a magnetic field aligned along an orthogonal direction
(say, $+z$). The total neutron energy is conserved, but due to
Zeeman splitting the two spin states in the direction of the $B$
field have different kinetic energies and travel with different
speeds. This is a situation typically encountered in the
so-called longitudinal Stern-Gerlach effect \cite{longSG} and in
neutron spin-echo experiments \cite{echo} (except that we are not
considering the second half of the evolution, with an opposite
$B$ field that recombines the two spin states). An experimental
realization of this situation was investigated very recently
\cite{BRSW}. If the initial wave function is
\andy{ondaspin}
\beq\label{eq:ondaspin}
|\Psi\rangle = |\psi\rangle\otimes |+\rangle_y = |\psi\rangle
\otimes\left(\frac{1}{\sqrt{2}}|+\rangle_z+\frac{i}
{\sqrt{2}}|-\rangle_z\right),
\eeq
where $|\pm \rangle_\alpha \;(\alpha=x,y,z)$ represents spin up/down
in direction $\alpha$, the final state in the position
representation, after crossing the $B$-field, reads
\andy{spinmag}
\beq\label{eq:spinmag}
\langle x|\Psi\rangle =
\frac{1}{\sqrt{2}}\;\psi\left(x+\frac{\Delta}{2}\right)\otimes
|+\rangle_z+\frac{i}{\sqrt{2}}\;\psi\left(x-\frac{\Delta}{2}\right)
\otimes |-\rangle_z.
\eeq
If only the $+y$-spin component is observed (``post selection" of the
initial spin component \cite{postsel}) the probability amplitude is
\andy{sypsi}
\beq
\label{eq:sypsi}
_y\langle +,x|\Psi\rangle =
\frac{1}{2}\left[\psi\left(x+\frac{\Delta}{2}\right)
+\psi\left(x-\frac{\Delta}{2}\right)\right],
\eeq
and the Wigner function is readily computed as
\andy{wigcats}
\barr
W^{\rm magn}(x,k,\Delta)&=&\frac{1}{4\pi}\exp[-2\delta^2(k-k_0)^2]
\left[\exp\left(-\frac{\left(x-x_0-\frac{\Delta}{2}\right)^2}{2\delta^2}\right)
\right. \nonumber\\ & & \left.
+\exp\left(-\frac{\left(x-x_0+\frac{\Delta}{2}\right)^2}{2\delta^2}\right)
+2
\exp\left(-\frac{(x-x_0)^2}{2\delta^2}\right)\cos(k\Delta)\right].
\label{eq:wigcats}
\earr
This result is slightly different from (\ref{eq:intcats}), because in
this case both spin components undergo a phase shift ($\pm
\Delta/2$).
Once again, for $\Delta=0$ (no magnetic field) one reobtains
(\ref{eq:wiggauss}).

We stress that in both cases the neutron wave packet has a
natural spread $\delta_t=\sqrt{\delta^2+(\hbar t/2m\delta)^2}$
(due to its free evolution for a time $t\simeq mL/\hbar k_0$);
however, this additional effect will be neglected, because, as
proved in Appendix A, it is not relevant for the loss of quantum
coherence.

\section{Fluctuating phase shift}
\label{sec-fluc}
\andy{fluc}
The previous analysis refers to a rather idealized case, in which
every neutron in the beam acquires a constant phase shift. This is
clearly not a realistic situation, for it does not take into
account the statistical fluctuations of the $B$ field or of the
shifter in the transverse section of the beam. If, for any reason,
the phase shift $\Delta$ fluctuates, the neutron beam will
partially loose its quantum coherence and the Wigner function
will be affected accordingly. We shall consider the case of
``slow" fluctuations, in the sense that each neutron crosses an
approximately static $B$ field (or a phase shifter of uniform
length $L$), but the intensity of the field (or the length of the
shifter) varies for different neutrons in the beam (different
``events"). We will suppose that every neutron undergoes a shift
$\Delta$ that is statistically distributed according to a
distribution law $w(\Delta)$. The collective ``degree of
disorder" of the shifts $\Delta$ can be given a quantitative
meaning in terms of the entropy
\andy{entr}
\beq\label{eq:entr}
S= -\int d\Delta\; w(\Delta)\; \log(w(\Delta)).
\eeq
On the other hand, the average Wigner function reads
\andy{wigmc}
\beq\label{eq:wigmc}
W_{\rm m}(x,k)= \int d\Delta\; w(\Delta)\; W(x,k,\Delta)
\eeq
and represents a partially mixed state. The coherence properties of
the neutron ensemble can be analyzed in terms of a {\em decoherence
parameter} \cite{FMP}
\andy{decpar}
\beq\label{eq:decpar}
\varepsilon =1-\frac{\mbox{Tr}\rho^2}{(\mbox{Tr}\rho)^2} =
1 - \frac{2\pi \int dx\; dk\; W_{\rm m}(x,k)^2} {\left(\int dx\;
dk\; W_{\rm m}(x,k)\right)^2} .
\eeq
This quantity measures the degree of ``purity" of a quantum state:
it is maximum when the state is maximally mixed
(Tr$\rho^2<$Tr$\rho$) and vanishes when the state is pure
(Tr$\rho^2=$Tr$\rho$): in the former case the fluctuations of $\Delta$
are large and the quantum mechanical coherence is completely lost,
while in the latter case $\Delta$ does not fluctuate and the
quantum mechanical coherence is perfectly preserved. The parameter
(\ref{eq:decpar}) was introduced within the framework of the
so-called ``many Hilbert space" theory of quantum measurements
\cite{NP1,NPN} and yields a quantitative estimate of decoherence.
The related quantity Tr$\rho-$Tr$\rho^2$ (that might be called
``idempotency defect") was first considered by Watanabe
\cite{Watanabe} many years ago. A measure of information for a quantum
system has been recently introduced, which is related to $\varepsilon$
and is more suitable than the Shannon entropy \cite{Bru}.

One might naively think that the two quantities $S$ and $\varepsilon$
should at least qualitatively agree: in other words, the loss of
quantum mechanical coherence should be larger when the neutron beam
interacts with fluctuating shifts of larger entropy. Such a naive
expectation turns out to be incorrect. Our purpose is to investigate
this problem. To this end, it is useful to consider some particular
cases.

\subsection{Gaussian noise}
\label{sec-wnoise}
\andy{wnoise}
We first assume that the shifts $\Delta$ fluctuate around their
average $\Delta_0$ according to a Gaussian law:
\andy{probdel}
\beq\label{eq:probdel}
w(\Delta) = \frac{1}{\sqrt{2\pi\sigma^2}}
\exp \left[-\frac{(\Delta-\Delta_0)^2}{2\sigma^2}\right],
\eeq
where $\sigma$ is the standard deviation. The ratio $\sigma/\Delta_0$
is simply equal to the ratio $\delta B/B_0$ (or $\delta L/L_0$),
$\delta B$ ($\delta L$) being the standard deviation of the
fluctuating magnetic field (length of phase shifter) and $B_0$
($L_0$) its average. The entropy of (\ref{eq:probdel}) is readily
computed from (\ref{eq:entr})
\andy{entwn}
\beq\label{eq:entwn}
S=\frac{1}{2}\log(2\pi e \sigma^2)
\eeq
and is obviously an increasing function of $\sigma$.

\subsubsection{Single Gaussian}
\begin{figure}
\begin{center}
\epsfig{file=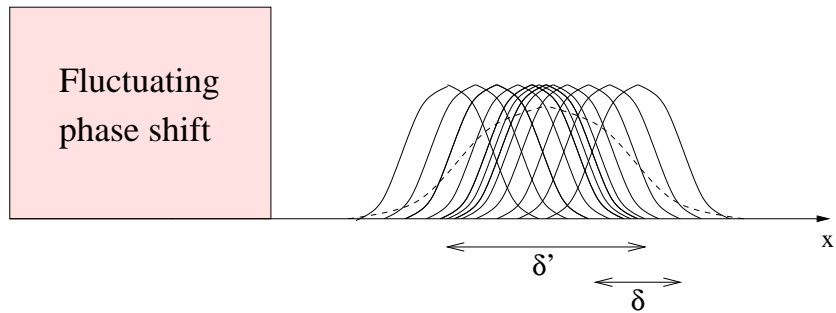,width=11cm}
\end{center}
\caption{If the phase shift fluctuates, each wave packet acquires
a different shift. This is pictorially represented in the figure,
where different outgoing wave packets are displayed, each
associated with a single neutron (``event"). The average Wigner
function is given by Eq.\ (\ref{eq:wmwn}).}
\label{fig:flucsh}
\end{figure}

Consider now a neutron described by a Gaussian wave packet. If the
phase shift $\Delta$ fluctuates according to (\ref{eq:probdel}),
the average Wigner function is readily computed by
(\ref{eq:wigmc}), (\ref{eq:wwigner}) and (\ref{eq:probdel}),
\andy{wmwn}
\beq\label{eq:wmwn}
W_{\rm
m}(x,k)=\frac{1}{\pi}\;\sqrt{\frac{\delta^2}{\delta^2+\sigma^2}}\;
\exp[-2\delta^2(k-k_0)^2]
\;\exp\left[-\frac{(x-x_0+\Delta_0)^2}{2(\delta^2+\sigma^2)}\right]
\eeq
and its marginals (\ref{eq:margx})-(\ref{eq:margp}) are
easily evaluated
\andy{marg1,marg2}
\barr
P(x) &=&
\frac{1}{\sqrt{2\pi(\delta^2+\sigma^2)}} \exp
\left[-\frac{(x-x_0+\Delta_0)^2}{2(\delta^2+\sigma^2)}\right] ,
\label{eq:marg1} \\
P(k) &=&
\sqrt{\frac{2\delta^2}{\pi} }
\exp \left[-2\delta^2(k-k_0)^2\right].
\label{eq:marg2}
\earr
Notice that the momentum distribution (\ref{eq:marg2}) is unaltered
and identical to $|\phi(k)|^2$ in (\ref{eq:gaussinv}): obviously, the
energy of each neutron does not change. Observe on the other hand the
additional spread in position $\delta'=\sqrt{\delta^2+\sigma^2}$
(Figure \ref{fig:flucsh}) and notice that the Wigner function and its
marginals are always normalized to one.

\begin{figure}
\begin{center}
\epsfig{file=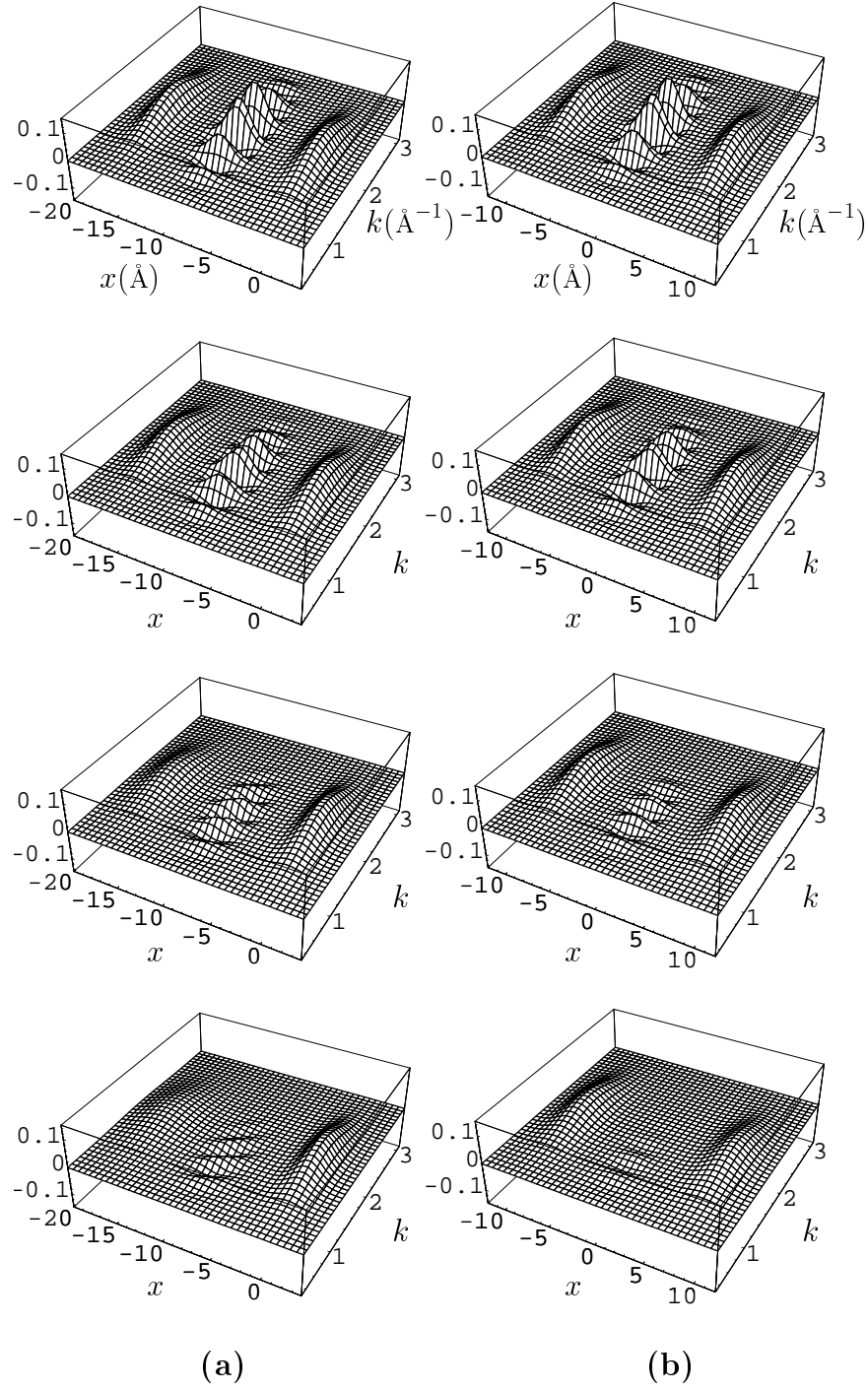,width=\textwidth}
\end{center}
\caption{Decoherence parameter vs coherence length of the wave
packet $\delta$(\AA) and standard deviation of the fluctuation
$\sigma$(\AA). (a) Gaussian wave packet. (b) Double Gaussian in an
interferometer, with $\Delta_0=16.1$\AA. (c) Double Gaussian in a
magnetic field, with $\Delta_0=16.1$\AA. In all cases
$k_0=1.7$\AA$^{-1}$. Observe that in case (a) the decoherence
parameter is a monotonic function of $\sigma$ for every value of
$\delta$, while this is not true for cases (b) and (c). Notice
also that in case (b) the decoherence parameter never reaches
unity ($\varepsilon\leq 3/4$): this is due to the fact that only
one Gaussian (in one branch of the interferometer) undergoes
statistical fluctuations [see Figure
\ref{fig:supp}(a)].}
\label{fig:decabc}
\end{figure}

The decoherence parameter (\ref{eq:decpar}) can be analytically
evaluated
\andy{decwn}
\beq\label{eq:decwn}
\varepsilon=1-\sqrt{\frac{\delta ^2}{\delta^2+\sigma^2}}
\eeq
and is a monotonic function of $\sigma$ for every value of $\delta$.
This behavior is in qualitative agreement with that of the entropy
(\ref{eq:entwn}). As expected, a more entropic distribution of phase shifts
entails a greater loss of quantum mechanical coherence for the
neutron ensemble. The behavior of $\epsilon$ vs $\delta$ and $\sigma$
is shown in Figure \ref{fig:decabc}(a).

\subsubsection{Double Gaussian in an interferometer}
\begin{figure}
\begin{center}
\epsfig{file=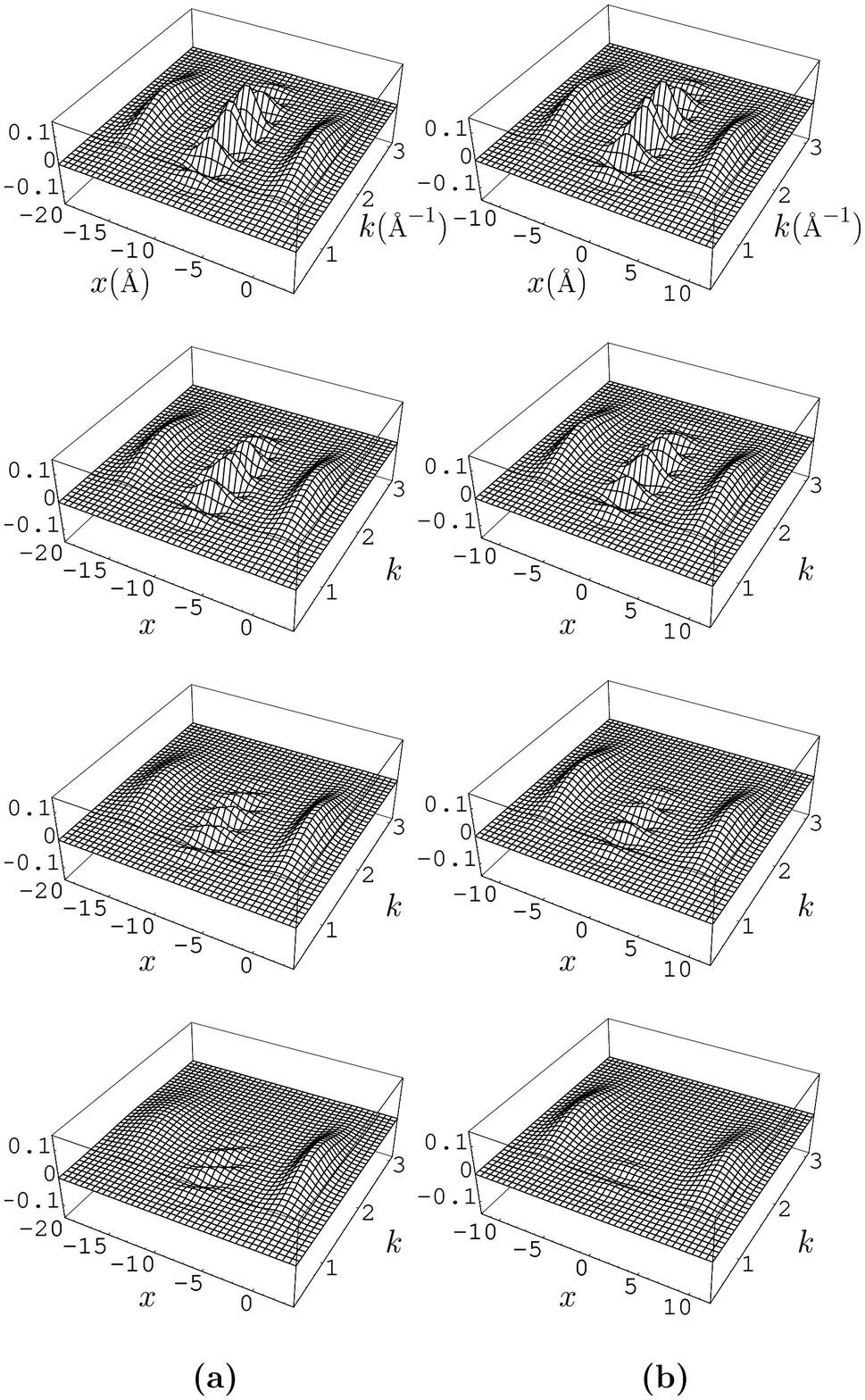,width=11.5cm}
\end{center}
\caption{Wigner functions for different
values of the standard deviation $\sigma$ in (\ref{eq:probdel}).
(a) Double Gaussian in an interferometer (\ref{eq:catsint}). (b)
Double Gaussian in a magnetic field (\ref{eq:wmcats}). From top
to bottom, $\sigma=0,0.6,1.2,1.8$\AA. The values of the other
parameters are $x_0=0$, $k_0=1.7$\AA$^{-1}$, $\delta=1.1$\AA,
$\Delta_0=16.1$\AA. Position $x$ and momentum $k$ are measured in
\AA$\;$and \AA$^{-1}$, respectively. Notice the
strong suppression of interference at large values of momentum,
both in (a) and (b). In case (a) only one of the two Gaussians
interacts with the fluctuating phase shifter; moreover, the
interference term in (\ref{eq:catsint}) depends on $x$ and the
oscillating part of the Wigner function is bent towards the
negative $x$-axis.}
\label{fig:supp}
\end{figure}

Consider now the double Gaussian state (\ref{eq:intcats}),
obtained when a neutron beam crosses an interferometer. The
average Wigner function (\ref{eq:wigmc}) reads
\andy{catsint}
\barr
& &W_{\rm m}^{\rm int}(x,k)=
\frac{\exp[-2\delta^2(k-k_0)^2]}{4\pi}\left\{\sqrt{\frac{\delta ^2}
{\delta^2+\sigma^2}}\exp\left[-\frac{(x+\Delta_0)^2}
{2(\delta^2+\sigma^2)}\right] + \exp\left[-\frac{x^2} {2\delta^2}
\right]\right.\nonumber\\ & &
\left.+2\sqrt{\frac{\delta ^2}
{\delta^2+\frac{\sigma^2}{4}}}
\exp\left[-\frac{\left(x+\frac{\Delta_0}{2}\right)^2+k^2\delta^2
\sigma^2}
{2\left(\delta^2+\frac{\sigma^2}{4}\right)}\right]
\cos\left(k\frac{2\delta^2\Delta_0-x\sigma^2}
{2\left(\delta^2+\frac{\sigma^2}{4}\right)}\right)\right\},
\label{eq:catsint}
\earr
where we set $x_0=0$ for simplicity. Its marginals
(\ref{eq:margx})-(\ref{eq:margp}) can both be computed
analytically; in particular, the momentum probability
distribution reads
\andy{margwn}
\beq\label{eq:margwn}
P(k)=\sqrt{\frac{\delta^2}{2\pi}}\exp\left[-2\delta^2(k-k_0)^2\right]
\left[1+\exp\left(-\frac{k^2\sigma^2}{2}\right)\cos(k\Delta_0)\right].
\eeq
As one can see from Figure \ref{fig:supp}(a), interference is
exponentially suppressed at high values of $k$ and the
oscillating part of the Wigner function is bent towards the
negative $x$-axis. This is due to the $x$-dependence of the
cosine term in (\ref{eq:catsint}) that entails different
frequencies for different values of $x$. The decoherence
parameter (\ref{eq:decpar}) reads
\andy{depaint}
\barr
\varepsilon&=&1-\frac{1}{4N^2}\left[\frac{1}{4}\left(1+\sqrt{\frac{\delta^2}
{\delta^2+\sigma^2}}\right)+\sqrt{\frac{\delta^2}
{4\delta^2+2\sigma^2}}\left(\exp\left[-\frac{\Delta_0^2}
{4\delta^2+2\sigma^2}\right]+\exp\left[-\frac{2k_0^2\delta^2\sigma^2}
{2\delta^2+\sigma^2}\right]\right)\right] \nonumber \\
&-&\frac{1}{2N^2}\left[\sqrt{\frac{\delta^2}
{4\delta^2+\sigma^2}}\exp\left[-\frac{\Delta_0^2+4k_0^2\delta^2\sigma^2}
{2(4\delta^2+\sigma^2)}\right]\cos\left[\frac{4k_0\Delta_0\delta^2}
{4\delta^2+\sigma^2}\right]\right. \nonumber \\
& &\left.+\frac{\delta^2}
{4\delta^2+\sigma^2}\exp\left[-\frac{\Delta_0^2+4k_0^2\delta^2\sigma^2}
{4\delta^2+\sigma^2}\right]\cos\left[\frac{8k_0\Delta_0\delta^2}
{4\delta^2+\sigma^2}\right]\right] \nonumber \\
&-&\frac{\delta^2}{N^2\sqrt{16\delta^4+12\delta^2\sigma^2+\sigma^4}}
\exp\left[\frac{(2\delta^2+\sigma^2)(\Delta_0^2+4k_0^2\delta^2\sigma^2)}
{16\delta^4+12\delta^2\sigma^2+\sigma^4}\right]\cos\left[
\frac{4k_0\Delta_0\delta^2(4\delta^2+3\sigma^2)}
{16\delta^4+12\delta^2\sigma^2+\sigma^4}\right],
\label{eq:depaint}
\earr
where the normalization
\barr
N &=& \int dx\; dk\; W_{\rm m}(x,k) \nonumber\\ &=& \frac{1}{2}
\left[1 +\sqrt{\frac{\delta^2}{\delta^2+\frac{\sigma^2}{4}}}
\exp\left(-\frac{\Delta_0^2+4\delta^2\sigma^2 k_0^2}
{8\left(\delta^2+\frac{\sigma^2}{4}\right)}\right)
\cos\left(\frac{\delta^2}{\delta^2+\frac{\sigma^2}{4}}
 k_0\Delta_0\right)\right]
\earr
represents the probability of detecting a neutron in the ordinary
channel. The explicit expression (\ref{eq:depaint}) 
of the decoherence parameter is
involved and difficult to understand. Therefore, $\varepsilon$ is
shown in Figure
\ref{fig:decabc}(b) as a function of $\delta$ and $\sigma$ for
fixed values of $k_0$ and $\Delta_0$: somewhat surprisingly, for
some values of $\delta$, even though the noise $\sigma$
increases, the decoherence $\varepsilon$ decreases.

Observe also that $\varepsilon$ never reaches unity: $\varepsilon
\leq 3/4$. This is due to the fact that one of the two Gaussians
does not undergo any fluctuations (there is a fluctuating phase
shifter in only one of the two routes of the interferometer):
therefore a part of the Wigner function is not affected by noise,
as one can see in Figure \ref{fig:supp}(a). We shall comment
again on the peculiar features of $\varepsilon$ in a while.

\subsubsection{Double Gaussian in a magnetic field}
If we consider a polarized neutron beam
interacting with a $B$ field perpendicular to its spin, Eq.\
(\ref{eq:wigcats}) yields
\andy{wmcats}
\barr
& &W_{\rm m}^{\rm magn}(x,k,\Delta)=
\frac{\exp[-2\delta^2(k-k_0)^2]}{4\pi}\left\{\sqrt{\frac{\delta ^2}
{\delta^2+\frac{\sigma^2}{4}}}\exp\left[-\frac{(x-\frac{\Delta_0}{2})^2}
{2(\delta^2+\frac{\sigma^2}{4})}\right]\right.\nonumber\\ & &
\left.\; + \; \sqrt{\frac{\delta ^2}
{\delta^2+\frac{\sigma^2}{4}}}\exp\left[-\frac{(x+\frac{\Delta_0}{2})^2}
{2(\delta^2+\frac{\sigma^2}{4})}
\right]+2\exp\left[-\frac{x^2}{2\delta^2}-\frac{k^2\sigma^2}{2}\right]
\cos(k\Delta_0)\right\},
\label{eq:wmcats}
\earr
This Wigner function has the same $k$-marginal (\ref{eq:margwn}) as
the previous one (although the $x$-marginals are different). Also in
this case, one observes a strong suppression of interference at large
values of momentum \cite{RS,BRSW,FMP}, but without $x$-dependence
in the cosine. See Figure
\ref{fig:supp}(b). In this case, the decoherence parameter
(\ref{eq:decpar}) reads
\andy{deccats}
\barr
\varepsilon&=&1-\frac{1}{4N^2}\;\sqrt{\frac{\delta ^2}{4\delta^2+\sigma^2}}\;\exp
\left(-\frac{\Delta_0^2+4k_0^2\delta^2\sigma^2}{4\delta^2+\sigma^2}\right)\;
\cos\left(\frac{8k_0\Delta_0\delta^2}{4\delta^2+\sigma^2}\right)\nonumber\\ &
&
-\frac{1}{4N^2}\;\sqrt{\frac{\delta ^2}{4\delta^2+\sigma^2}}\;\left[1+\exp\left(
-\frac{\Delta_0^2}{4\delta^2+\sigma^2}\right)+\exp\left(-\frac{4k_0^2\delta^2
\sigma^2}{4\delta^2+\sigma^2}\right)\right]\nonumber\\ & &
-\frac{4}{N^2}\;
\frac{\delta^2}{8\delta^2+\sigma^2}\;\exp
\left(-\frac{\Delta_0^2+4k_0^2\delta^2\sigma^2}{8\delta^2+\sigma^2}\right)\;\cos\left(
\frac{8k_0\Delta_0\delta^2}{8\delta^2+\sigma^2}\right).
\label{eq:deccats}
\earr
Again, the explicit expression of the decoherence parameter is
complicated and depends on several physical parameters; it is
therefore convenient to concentrate on a particular case. An
experimental realization of a fluctuating shift (according to a
given statistical law) is easier with the magnetic field
arrangement discussed in Section
\ref{sec-dGauss}. Let us therefore consider the experiment
\cite{BRSW}, in which a polarized ($+y$) neutron enters a magnetic
field, perpendicular to its spin, of intensity $B_0=0.28$mT, confined
in a region of length $L=57$cm. The average neutron wavenumber is
$k_0=1.7
\cdot 10^{10}$m$^{-1}$ and its coherence length (defined by a
chopper) is $\delta=1.1\cdot 10^{-10}$m. By travelling in the
magnetic field, the two neutron spin states are separated by a
distance $\Delta_0=2 m\mu B_0 L/\hbar^2 k_0^2= 16.1
\cdot 10^{-10}$m, one order of magnitude larger than $\delta$.
The behavior of $\varepsilon$ in (\ref{eq:deccats}) is shown in
Figure 2(c) for these experimental values: observe that for
$\delta\gtrsim 3$\AA\, $\epsilon$ is {\em not} a monotonic
function of $\sigma$: in other words, for some values of the
parameters, even though the noise $\sigma$ increases, the
decoherence $\epsilon$ decreases. This is at variance with the
behavior of the entropy (\ref{eq:entwn}) and with what one might
naively expect. We conclude that, in general, both for a double
Gaussian in an interferometer and in a magnetic field, the
behavior of $\varepsilon$ does not agree with that of the entropy.

\subsection{Sinusoidal fluctuations with increasingly less rational
frequencies}
\label{sec-fib}
\andy{fib}
In order to shed some more light on the results of the previous
subsection, let us consider a different example, that is more
convenient for an experimental perspective. Suppose that the phase
shift changes according to the law
\andy{Dcos1}
\beq
\Delta(t)=\Delta_0+\Delta_1[\sin(\Omega t)+\sin(r_j\Omega t)].
\label{eq:Dcos1}
\eeq
where $t$ is time, $\Omega$ a frequency much smaller than $v_0/L$,
the inverse time of flight of the neutron in the shifter's region,
$\Delta_0$ the mean phase shift, $\Delta_1 (\ll \Delta_0)$ the
``fluctuation" width (see below) and $r_{j} \; (0<r_j<1)$ a real
number. For the neutron ensemble (the beam) the shifts will be
distributed according to law
\andy{probcos}
\beq\label{eq:probcos}
w(\Delta)=\int dt\;f(t)\;\delta(\Delta-\Delta(t)),
\eeq
where $f(t)$ is the probability density function of the stochastic
variable $t$. In our case, $f(t)=1/T$ in $(0,T)$, where $T(\gg
\Omega^{-1})$ is a sufficiently large time interval.
In such a case, by making use of (\ref{eq:probcos}), the Wigner
function can be expressed as an ergodic average
\andy{wicos1}
\beq
 W_{\rm m}(x,k)=\int d\Delta\; w(\Delta)\; W(x,k,\Delta)=
\frac{1}{T}\int_T dt\;W(x,k,\Delta(t)).
\quad (T \; \mbox{large})
 \label{eq:wicos1}
\eeq
We stress that $\Delta$ is treated like a random variable
although, strictly speaking, the underlying process is
deterministic. However, this is not a conceptual difficulty: in
practice, one just treats the neutron ensemble in an experimental
run without looking at the correlations among different neutrons.
The same effects on the neutron ensemble would be obtained by
first generating a random variable $t$, uniformly distributed in
$(0,T)$, then constructing the additional random variable
$\Delta$ according to (\ref{eq:Dcos1}) and finally accumulating
all neutrons in the experimental run. In this way different
neutrons are uncorrelated. The distribution law of the shifts
(\ref{eq:Dcos1}) can be obtained by means of a $B$-field
\andy{fibfie}
\beq\label{eq:fibfie}
B(t)=B_0+B_1\left[\sin(\Omega t)+\sin(r_j \Omega t)\right].
\eeq
Like in the previous section, we assume that $B$ is a slowly
varying function of time, so that each neutron experiences a
static field during its interaction. Observe that the scheme
proposed in (\ref{eq:Dcos1})-(\ref{eq:wicos1}) is not difficult
to realize experimentally. On the other hand, it would be
complicated to obtain the same distribution of shifts with a phase
shifter placed in one of the two routes of an interferometer.

We will study the coherence properties of the neutron beam when it
crosses a magnetic field made up of two ``increasingly less
rational" frequencies, by choosing
\andy{fibrat}
\beq\label{eq:fibrat}
r_j=\frac{f_j}{f_{j+1}},
\eeq
where $f_j$ are the Fibonacci numbers
\andy{fib}
\beq\label{eq:fib}
f_{j+1}=f_{j}+f_{j-1} \qquad (f_0=f_1=1).
\eeq
This particular choice is motivated by the (naive) expectation
that an oscillating magnetic field (\ref{eq:fibfie}) composed of
mutually less rational frequencies should provoke more
decoherence on the neutron ensemble. Once again, this expectation
will turn out to be incorrect. The ratios (\ref{eq:fibrat}) tend
to the golden mean (the ``most irrational" number
\cite{ODA}) as $j$ increases
\andy{fibgr}
\beq\label{eq:fibgr}
r_{j}\stackrel{j\to\infty}{\longrightarrow} r_\infty =
\frac{\sqrt{5}-1}{2}.
\eeq
In general, one cannot obtain an analytic expression for the
probability density function (\ref{eq:probcos}); however, an
accurate numerical evaluation of $w(\Delta)$ is possible: for
every finite value of $j$, $r_j$ is a rational number, so that
one can integrate (\ref{eq:probcos}) over the interval 
$T=f_{j+1}2\pi/\Omega$. In Figure
\ref{fig:resfib} we show the results of our numerical analysis. 
The distribution function $w(\Delta)$ has a finite number of
(integrable) divergences in its interval of definition; as the
order in the Fibonacci sequence becomes higher, the number of
divergences in the interval grows. In the $j\to\infty$ limit, i.e.
for the golden mean $r_\infty=(\sqrt{5}-1)/2$, it is possible to
apply the theorem on averages for the ergodic motion on a torus
\cite{Arnold} and find an analytical expression of $w(\Delta)$ in
terms of an elliptic integral of first kind (see
Appendix~B). The
resulting distribution is a smooth function with only one
(integrable) divergence in $\Delta=0$ and is plotted in
Fig.~\ref{fig:resfib}(f).

By applying the same technique utilized for 
the numerical evaluation of $w$, the entropy is
computed according to the formula
\andy{entcos}
\beq\label{eq:entcos}
S= -\int d\Delta\; w(\Delta)\; \log(w(\Delta)) =
- \frac{1}{T}\int_T dt\;\log[w(\Delta(t))],
\eeq
which is easily obtained by Eqs.\ (\ref{eq:entr})
and (\ref{eq:probcos}) (using
the value $T=f_{j+1}2\pi/\Omega$ for the numerical evaluation).

The decoherence parameter is computed from Eq.\ (\ref{eq:decpar}),
first with the Wigner function (\ref{eq:wwigner}) (single Gaussian)
and then with the Wigner function (\ref{eq:wigcats}) (double Gaussian
in a magnetic field): in both formulas, we used Eq.\
(\ref{eq:wicos1}) and set $\Delta_0=16.1\cdot10^{-10}$m,
$\Delta_1=2\cdot10^{-10}$m and the same numerical values of the
previous subsection for $k_0$ and $\delta$ \cite{BRSW}. Our results
are summarized in Table 1 and Figure \ref{fig:graph}.

We notice that, although, for $j=1,...,5$, $S$ is 
a monotonically increasing
function of the Fibonacci number in the sequence, $\varepsilon$
reaches a maximum for $r_j=3/5$ (i.e., $j=3$). It is remarkable
that the maximum is obtained for the same Fibonacci ratio in both
cases (single and double Gaussian). Once again, the behavior of
entropy and decoherence are qualitatively different. Figure
\ref{fig:graph} should be compared to Figure \ref{fig:decabc}: it
is worth noting that in the case analyzed in this section, unlike
in Section \ref{sec-wnoise}, the behavior of entropy and
decoherence do not agree even when the neutron state is a single
Gaussian (namely, a ``classical" state).

\section{Conclusions}
\label{sec-conc}
\andy{conc}
Decoherence is a very useful concept, that has recently been
widely investigated and has turned out to be very prolific. It is
intuitively related to the loss of ``purity" of a quantum
mechanical state and can be given a quantitative definition, as in
(\ref{eq:decpar}). However, we have seen that the very notion of
decoherence is delicate: in particular, it is not correct to think
that a quantum system, by interacting with an increasingly
``disordered" environment, will suffer an increasing loss of
quantum coherence. Our analysis has been performed by assuming
that each neutron, during an experimental run, interacts with a
constant magnetic field: the neutron beam, on the average,
undegoes decoherence. This ``quasistatic" approximation is only a
working hypothesis and will be relaxed in future work. As
emphasized at the beginning of Sec.\ \ref{sec-fib}, it is easy to
achieve experimentally in the limit $\Omega \ll v_0/L$, where
$\Omega$ is a characteristic frequency of the fluctuation and
$(v_0/L)^{-1}$ the time of flight of the neutron in the phase
shifter or in the magnetic field. This approximation also
simplifies (both conceptually and technically) our theoretical
analysis, without however having a substantial influence on our
general conclusions. It is worth stressing, in this respect, that
the decoherence parameter, defined in (\ref{eq:decpar}), depends
on the interaction and not on the free Hamiltonian, at least in
the physical situations investigated here (see Appendix A).

Our quantitative definition of decoherence depends, as it should, on
the very characteristics of the experimental setup: the decoherence
parameter is defined in terms of the average Wigner function (or
equivalently the density matrix) of the neutron ensemble, after the
interaction with the apparatus. An experimental check of the features
of the Wigner functions
discussed in this paper would require its tomographic observation.
Similar techniques are commonly
applied in quantum optics
\cite{qtom} and would be available in neutron optics as well, in
particular for the experimental arrangement discussed in Section
\ref{sec-fib}. However, we think that a better
comprehension of the effects analyzed in this paper could probably be
achieved by studying the marginals (or possibly some other
tomographic projection) of the Wigner function and the visibility of
the interference pattern. Additional work is in progress in this
direction. From an experimental perspective, an analysis of
decoherence effects along the guidelines discussed here would be
challenging: although the concepts of decoherence and entropy are
intuitively related, they display some interesting differences. If
properly understood, those situations in which a larger noise yields
a more coherent quantum ensemble might lead to unexpected
applications.

\section*{Acknowledgements}
We thank H. Rauch and M. Suda for useful comments and I. Guarneri for
an interesting remark. The numerical computation was performed on the
``Condor" pool of the Italian Istituto Nazionale di Fisica Nucleare
(INFN). This work was realized within the framework of the TMR
European Network on ``Perfect Crystal Neutron Optics"
(ERB-FMRX-CT96-0057).

%%%%%%%%%%%%%%%%%%%%%%%%%%%%%%%%%%%%%%%%%%%%%%%%%%
%%%%%%%%%%%%%% Automatic numbering A, B, C ...
%\appendix
%\setcounter{section}{0}
%\setcounter{equation}{0}
%\section{Appendix A} -------> A APPENDIX A
%%%%%%%%%%%%%%%%%%%%%%%%%%%%%%%%%%%%%%%%%%%%%%%%%%
%%%%%%%%%%%%%%%%%%%%%%%%%%%%%%%%%%%%%%%%%%%%%%%%%%
%\setcounter{equation}{0}
%\section{Appendix B} -------> B APPENDIX B
%%%%%%%%%%%%%%%%%%%%%%%%%%%%%%%%%%%%%%%%%%%%%%%%%%

\renewcommand{\thesection}{\Alph{section}}
\setcounter{section}{1}
\setcounter{equation}{0}
\section*{Appendix A}
\label{sec-appA}
\andy{appA}

\renewcommand{\thesection}{\Alph{section}}
\renewcommand{\thesubsection}{{\it\Alph{section}.\arabic{subsection}}}
\renewcommand{\theequation}{\thesection.\arabic{equation}}

We prove that the decoherence parameter (\ref{eq:decpar}), under
rather general conditions, does not depend on the free evolution
of a quantum system. Let the Hamiltonian of a quantum system be
\andy{ham1}
\beq
\label{eq:ham1}
H=H_0+H_1(\alpha),
\eeq
where $H_0$ and $H_1$ are the free and interaction Hamiltonians,
respectively, and $\alpha$ is a c-number (that can fluctuate
according to a given statistical law). We assume that
\andy{com}
\beq
\label{eq:com}
[H_0,H_1(\alpha)]=i C, \qquad \mbox{with}\qquad
[H_0,C]=[H_1(\alpha),C]=0,
\eeq
where $C$ is in general an operator (independent of $\alpha$), so
that ($\hbar=1$)
\andy{com1}
\beq
\label{eq:com1}
e^{-it (H_0+H_1(\alpha))}=e^{i t^2 C/2} e^{-it H_0} e^{-it H_1(\alpha)}.
\eeq
Consider now the density matrix at time $t$,
\andy{rotalp}
\beq
\label{eq:rotalp}
\rho_\alpha(t)=e^{-it (H_0+H_1(\alpha))}\;\rho_0
\;e^{it(H_0+H_1(\alpha))},
\eeq
where $\rho_0$ is the initial density matrix. From Eq.\
(\ref{eq:com1})
\andy{rotcom}
\beq
\label{eq:rotcom}
\rho_\alpha(t)=e^{i t^2 C/2}\;e^{-it H_0}\;e^{-it
H_1(\alpha)}\;\rho_0\; e^{it  H_1(\alpha)}\;e^{itH_0}e^{-i t^2C/2}
\eeq
and the average over $\alpha$ yields
\andy{rotm}
\beq\label{eq:rotm}
\overline{\rho(t)}=e^{i t^2C/2}\;e^{-it H_0}\left( \int d\alpha \;
w(\alpha)\;{\rm e}^{-it H_1(\alpha)}\;\rho_0\; {\rm e}^{it
H_1(\alpha)}\right)e^{itH_0}e^{-i t^2C/2},
\eeq
where $w(\alpha)$ is the distribution function and the bar
denotes average. Therefore
\andy{theo}
\beq\label{eq:theo}
{\rm Tr}\left[\overline{\rho(t)}\right]= {\rm Tr}
\left[\overline{\rho_{\rm int}(t)}\right],
\eeq
where $\rho_{\rm int}$ is the density matrix in the following
interaction picture:
\andy{intpi}
\beq\label{eq:intpi}
\rho_{\rm int}(t)
=e^{itH_0-it^2 C/2}\;\rho(t)
\;e^{-itH_0+it^2 C/2}
=e^{-itH_1(\alpha)}\;\rho_0\;e^{itH_1(\alpha)}.
\eeq
This proves that the trace of the average density matrix does not
depend on the free evolution. The result (\ref{eq:theo}) can be
generalized to any function of the average density matrix
\andy{trthe}
\beq\label{eq:trthe}
{\rm Tr}\left[f\left(\overline{\rho(t)}\right)\right]={\rm Tr}
\left[f\left(\overline{\rho_{\rm int}(t)}\right)\right].
\eeq
This shows that the decoherence parameter defined in
(\ref{eq:decpar}) does not depend on the free evolution:
\andy{finthe}
\beq\label{eq:finthe}
\varepsilon=\varepsilon_{\rm int},
\eeq
as claimed at the end of Sec.\ \ref{sec-prel}. This result can be
applied to the case studied in Sec.\ \ref{sec-fluc}, where
$H_0=\frac{p^2}{2m}, H_1=-\bm \mu \cdot \bm B$ and the parameter
$\alpha$ is the intensity of the magnetic field $\bm B$ (whose
direction is supposed constant). Notice that we are considering
wave packets that interact with a constant and homogeneous field
$\bm B$ from the initial time $t=0$ to the final time $t\simeq
mL/\hbar k_0$, so that condition (\ref{eq:com}) is fulfilled. The
case of a neutron wave packet in an interferometer is analogous,
if we assume that the phase shifter simply yields a phase
(optical potential approximation).

\setcounter{section}{2}
\setcounter{equation}{0}
\section*{Appendix B}
\label{sec-appB}
\andy{appB}

\begin{figure}
\begin{center}
\epsfig{file=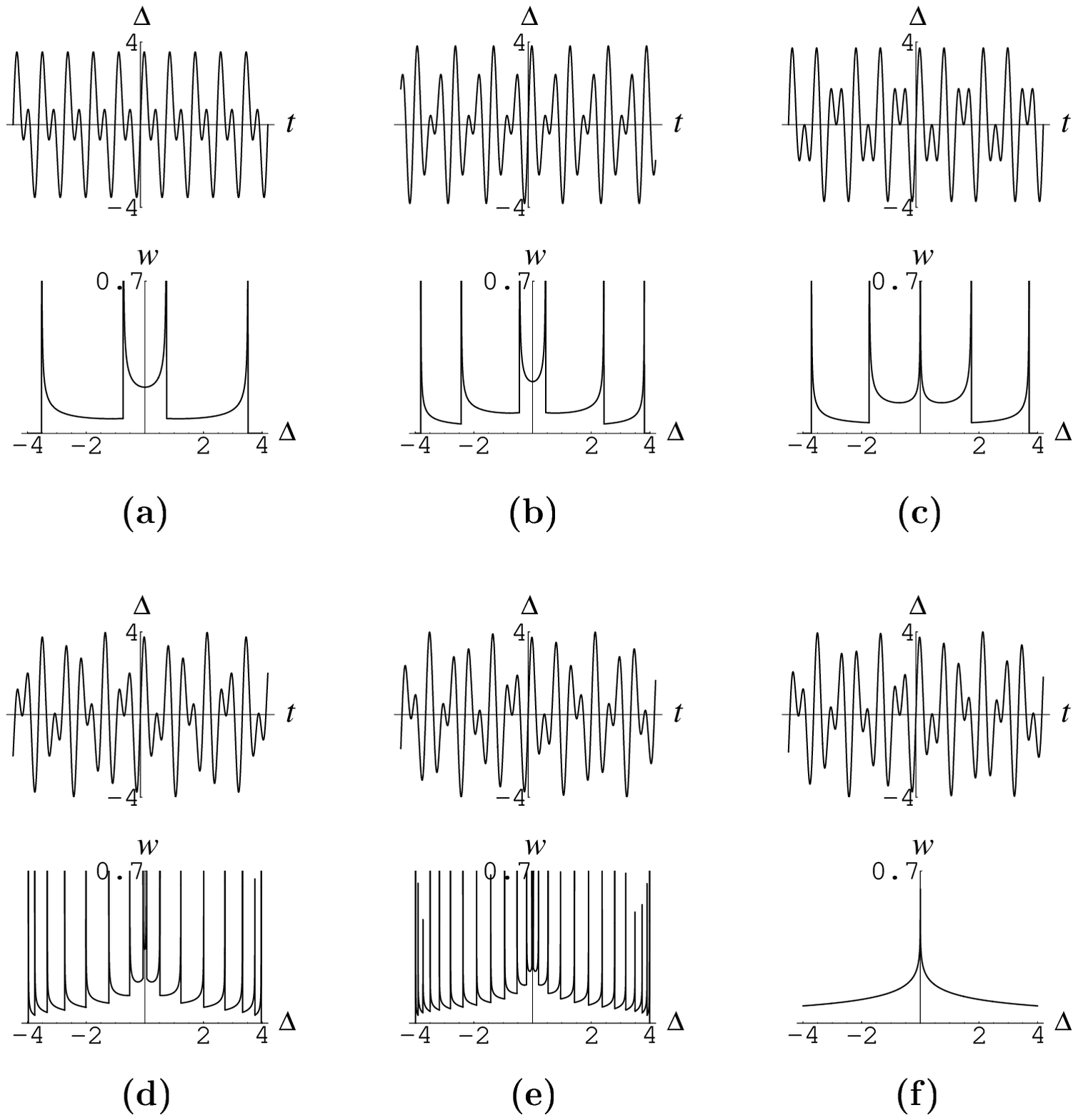,width=\textwidth}
\end{center}
\caption{Phase shift $\Delta$ in (\ref{eq:Dcos1}) and distribution
function $w(\Delta)$ in (\ref{eq:probcos}), for different values
of $r_j$: (a) $r_1=1/2$; (b) $r_2=2/3;$ (c) $r_3=3/5;$ (d)
$r_4=5/8;$ (e) $r_5=8/13;$ (f) $r_\infty=(\protect\sqrt{5}-1)/2$.
In each figure $\Delta_1=2$ (we set $\Delta_0=0$ for clarity of
presentation): above, phase shift $\Delta(t)$; below,
distribution function $w(\Delta)$. Notice that, by increasing $j$
(index of the Fibonacci sequence), the two frequencies become
mutually ``less rational," the phase shift $\Delta(t)$ becomes
more irregular and its distribution function $w(\Delta)$ more
uniform. [The entropy behaves accordingly, increasing for
$j=1,\ldots, 5$ (see Fig.\ \ref{fig:graph}(a))]. Notice that the
number of divergences of the distribution function increases with
$j$; as shown in Appendix B, in the $j=\infty$ limit, the
distribution becomes continuous with only one (logarithmic)
divergence in $\Delta=0$ and can be expressed as an elliptic
integral (\ref{eq:paofin}).}
\label{fig:resfib}
\end{figure}

\begin{figure}
\begin{center}
\epsfig{file=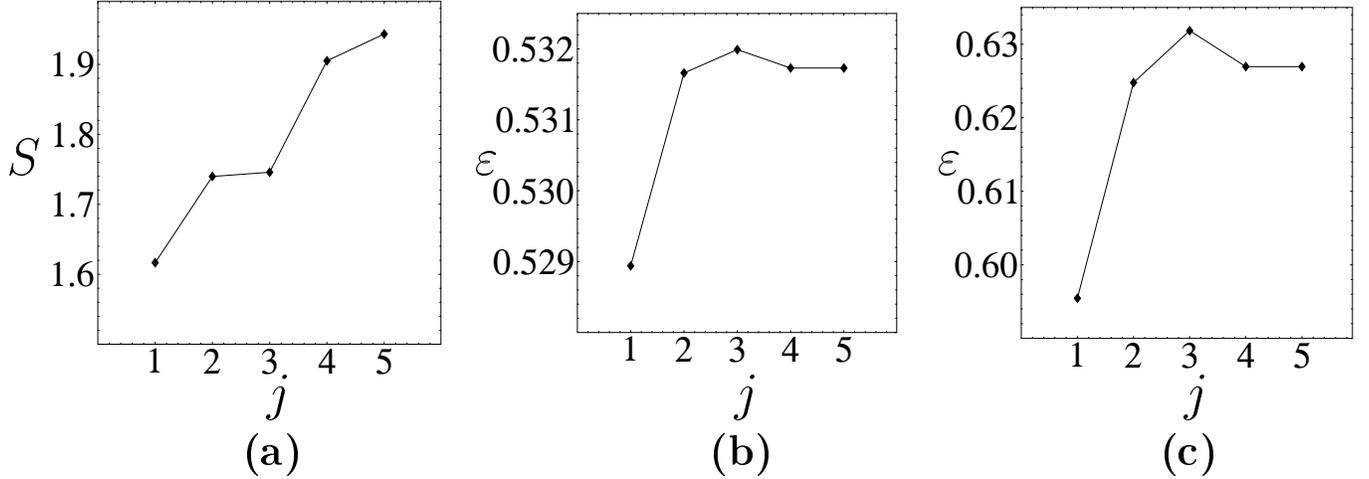,width=\textwidth}
\end{center}
\caption{(a) Entropy (\ref{eq:entcos}) vs $j$
(index in the Fibonacci sequence). (b) Decoherence parameter
(\ref{eq:decpar}) vs $j$: case of a single Gaussian. (c)
Decoherence parameter (\ref{eq:decpar}) vs $j$: case of a double
Gaussian in a magnetic field. Notice that, while the entropy is an
increasing function of $j$ for $j=1,\ldots, 5$, the decoherence
parameter displays a maximum at $j=3$, both for a single and a
double Gaussian.}
\label{fig:graph}
\end{figure}

We compute here the distribution function (\ref{eq:probcos}) when
$f(t) = 1/T$ and $\Delta(t)$ changes according to (\ref{eq:Dcos1}),
in the $j=\infty$ limit (\ref{eq:fibgr}). The function $w(\Delta)$
has a finite number of (integrable) divergences in its interval of
definition. Notice that, as the order in the Fibonacci sequence
becomes higher, the number of divergences in the interval grows and
the numerical evaluation of the distribution function becomes more
difficult. Let us introduce the two-component vector
\andy{evo2co}
\beq\label{eq:evo2co}
\bm\varphi_j(t)=(\varphi_1,\;\varphi_2)=\bm\omega_j t, \quad \mbox{where}
\quad \bm\omega_j=(\Omega,\; r_j\Omega).
\eeq
The vector $\bm\varphi_j$ performs a (quasi)periodic motion on the
two-dimensional torus $T^2$. In particular, for every {\em finite}
value of $j$ the frequencies are dependent, i.e.
$\omega_2/\omega_1=r_j\in \mathbb{Z}$, and the orbits are closed. For
larger values of $j$ the number of windings in a period increases and
the length of the periodic orbit becomes larger. In the $j\to\infty$
limit, the two frequencies become independent and the resulting
motion on the 2-torus becomes ergodic: the trajectory is everywhere
dense and uniformly distributed on $T^2$. In this case, according to
the theorem on averages
\cite{Arnold}, the time average of every integrable function $f(\bm\varphi)$
(where $\bm\varphi \equiv \bm\varphi_\infty$) coincides with its
space average, i.e.\
\andy{meantheo}
\beq
\lim_{T\to\infty}\frac{1}{T}\int_0^T dt\;f(\bm\varphi(t))=
\frac{1}{(2\pi)^2}\int_0^{2\pi}d\varphi_1\int_0^{2\pi}d\varphi_2\;
f(\bm\varphi).
\label{eq:meantheo}
\eeq
Applying Eq.~(\ref{eq:meantheo}) to the function
\andy{ff}
\beq\label{eq:ff}
f(\bm\varphi)=\delta(\Delta-\Delta_1[\sin\varphi_1 +\sin\varphi_2])
\eeq
we obtain
\andy{ergopao}
\barr\label{eq:ergopao}
w(\Delta)&=&\lim_{T\to\infty}\frac{1}{T}\int_0^T\;dt\;
\delta(\Delta-\Delta_1[\sin(\Omega t)+\sin(r_\infty\Omega t]))
\nonumber \\
&=&\frac{1}{(2\pi)^2}\int_0^{2\pi}d\varphi_1\int_0^{2\pi}d\varphi_2\;
\delta(\Delta-\Delta_1[\sin\varphi_1 +\sin\varphi_2])\nonumber\\
%&=&\int_{-1}^1\;ds_1\;P_S(s_1)\;
%\int_{-1}^1\;ds_2\;P_S(s_2)\;\delta(\Delta-\Delta_1s_1-\Delta_1s_2),
&=&\frac{1}{\Delta_1}\int_{-1}^{+1} ds\;P_S(s)\;
P_S( \Delta/\Delta_1-s ),
\earr
where $P_S$ is the sine distribution
\andy{senpao}
\beq\label{eq:senpao}
P_S(s)=\frac{1}{2\pi}
\int_0
^{2\pi} d\varphi\;\delta(s-\sin\varphi)=
\frac{1}{\pi\sqrt{1-s^2}}.
\eeq
After some algebraic manipulation one finds
\andy{paofin}
\beq\label{eq:paofin}
w(\Delta)=\frac{2}{\pi^2\Delta_1}\;
F\left(\arcsin\frac{1}{\sqrt{1+\frac{|\Delta|}{2\Delta_1}}},
\sqrt{1-\left(\frac{\Delta}{2\Delta_1}\right)^2}\right),
\eeq
where $F(\beta,\gamma)$ is the elliptic integral of first kind
\cite{GR}
\andy{ptelli}
\beq\label{eq:ptelli}
F(\beta,\gamma)=\int_0^\beta\;d\alpha\;
\frac{1}{\sqrt{1-\gamma^2\sin^2\alpha}}.
\eeq
The limiting distribution function (\ref{eq:paofin}) is plotted in
Fig.~\ref{fig:resfib}(f).

Observe that in Fig.~\ref{fig:resfib}(a-f) the number of divergences
increases so quickly that, in the $j=\infty$ limit (golden mean),
$w(\Delta)$ becomes a smooth function with only one (integrable)
divergence in $\Delta=0$ (indeed $w(\Delta)\sim
\log(8\Delta_1/|\Delta|)/\pi^2\Delta_1$ for $\Delta\to0$).
In this sense Berry {\em et al.} coined the epigram ``stocasticity
is the ubiquity of catastrophe" \cite{Berry}. (Incidentally,
notice the similarity of Fig.~\ref{fig:resfib} with Fig.\ 12 of
\cite{Berry}.)

%%%%%%%%%%%%%%%%%%%%%%% REFERENCES %%%%%%%%%%%%%%%%%%%%%%%%%%%%%%%

%
%
%
%%%%%%%%%%figures%%%%%%%%%%%%%%%%%%
\begin{table}
\caption{Entropy and Decoherence \label{fibo}}
\begin{center}
\begin{tabular}{c|c|c|c|c|}
$j$ & $r_j$ & $S$ & $\varepsilon$ (single Gaussian) & $\varepsilon $
(double Gaussian)
\\\hline
1& 1/2 &1.6165 &0.52894 &0.59545
\\ 2& 2/3 &1.7398 &0.53166 &0.62478 \\ 3& 3/5 &1.7458 &0.53199 &0.63184
\\ 4& 5/8 &1.9051&0.53173 &0.62695 \\ 5& 8/13 &1.9434 &0.53173 &0.62695
\\
\end{tabular}
\end{center}
\end{table}


\begin{thebibliography}{99}

\bibitem{Dec} \andy{Dec}
D. Giulini {\em et al}, {\it Decoherence and the Appearance of a Classical
World in Quantum Theory} (Springer, Berlin, 1996).

\bibitem{NPN} \andy{NPN}
M. Namiki, S. Pascazio and H. Nakazato, {\it Decoherence and Quantum
Measurements} (World Scientific, Singapore, 1997).

\bibitem{neutron} \andy{neutron}
H. Bonse and H. Rauch, eds., {\it Neutron Interferometry} (Clarendon,
Oxford, 1979); G. Badurek, H. Rauch and A. Zeilinger, {\it Matter
Wave Interferometry} (North-Holland, Amsterdam, 1988); H.
Rauch and S. A. Werner, {\it Neutron Interferometry: Lessons
in Experimental Quantum
Mechanics} (Oxford University Press, Oxford, 2000).

\bibitem{longSG} \andy{longSG}
F. Mezei, Physica {\bf B151}, 74 (1988); R. Golub, R. G\"ahler and T.
Keller, Am. J. Phys. {\bf 62}, 779 (1994).

\bibitem{echo} \andy{echo}
F. Mezei, Z. Phys. {\bf 25}, 146 (1972); {\it Neutron Spin Echo},
Lecture Notes in Physics {\bf 128} (Springer Verlag, Berlin, 1980).

\bibitem{Wigner} \andy{Wigner}
E. Wigner, Phys. Rev. {\bf 40}, 749 (1932); M. Hillery {\em et al},
Phys. Rep. {\bf 106}, 121 (1984).

\bibitem{QOpt} \andy{QOpt}
R. J. Glauber, Phys. Rep. {\bf 130}, 2766 (1963).

\bibitem{NP1} \andy{NP1}
M. Namiki and S. Pascazio, Phys. Lett. {\bf A147}, 430 (1990); Phys.
Rev. {\bf A44}, 39 (1991).

\bibitem{RS} \andy{RS}
H. Rauch and M. Suda, Physica {\bf B241-243}, 157 (1998); J. Appl.
Phys. {\bf B60}, 181 (1995); H. Rauch, M. Suda and S. Pascazio,
Physica {\bf B267}, 277 (1999).

\bibitem{Saito} \andy{Saito}
N. Saito in {\it Quantum Physics, Chaos Theory and Cosmology}, M.
Namiki ed. (AIP, New York, 1996) p.\ 275; M. Namiki (private
communication); H. Nakazato {\em et al}, Phys. Lett. {\bf A222},
130 (1996); N. Saito and H. Makino, {\it Quantum chaos,
ergodicity, and quantum measurement. General considerations},
preprint Waseda University, Japan (2000).

\bibitem{BRSW} \andy{BRSW}
G. Badurek {\em et al}, Optics Comm. {\bf 179}, 13 (2000).

\bibitem{postsel} \andy{postsel}
H. Rauch {\em et al}, Phys. Rev. {\bf A53}, 902 (1996).

\bibitem{FMP} \andy{FMP}
P. Facchi, A. Mariano and S. Pascazio, Acta Phys. Slov. {\bf 49},
677 (1999); Physica {\bf B276-278}, 970 (2000).

\bibitem{Watanabe} \andy{Watanabe}
S. Watanabe, Z. Phys. {\bf 113}, 482 (1939).

\bibitem{Bru} \andy{Bru}
\v C. Brukner and A. Zeilinger, Phys. Rev. Lett. {\bf 83}, 3354 (1999).

\bibitem{ODA} \andy{ODA}
A. M. Ozorio de Almeida, {\it Hamiltonian Systems: Chaos and
Quantization} (Cambridge University Press, Cambridge, 1988), Section 5.3.

\bibitem{Arnold} \andy{Arnold}
V. I. Arnol'd, {\it Mathematical methods of classical mechanics}
(Springer Verlag, Berlin, 1989), \S 51.

\bibitem{qtom} \andy{qtom}
W. Vogel and D. G. Welsh, {\it Lectures on Quantum Optics} (Akademie
Verlag/VCH Publishers, Berlin, 1994); W. P. Schleich, M. Pernigo and F.
Le Kien, Phys. Rev. {\bf A44} 2172 (1991); G. Breitenbach,
S. Schiller and J. Mlynek, Nature {\bf 387}, 471 (1997); D. Vitali, P.
Tombesi and G. J. Milburn, Phys. Rev. {\bf A57}, 4930 (1998).

\bibitem{GR} \andy{GR}
I.\ S. Gradshteyn and I.\ M. Ryzhik, {\em Table of Integrals, Series,
and Products} (Academic Press, San Diego, 1994),
\S 8.111.

\bibitem{Berry} \andy{Berry}
M. V. Berry, N. L. Balazs, M. Tabor and A. Voros, Ann. Phys. {\bf
122}, 26 (1979).

\end{thebibliography}
 \end{document}